\documentstyle[11pt,newpasp,twoside,epsf]{article}
\def\edcomment#1{\iffalse\marginpar{\raggedright\sl#1\/}\else\relax\fi}
\reversemarginpar

\begin{document}
\title{Gamma-Ray Burst Luminosity Functions Based on Luminosity Criteria}
 \author{Maarten Schmidt}
\affil{California Institute of Technology, Pasadena, CA 91125, USA}

\begin{abstract}

We review five studies of the gamma-ray burst (GRB) luminosity function 
that are based on one or more of the proposed luminosity criteria, i.e., 
variability, spectral lag and spectral hardness. A comparison between
the resulting luminosity functions at redshift zero shows considerable
divergence. We derive for each of the luminosity functions the 
expected source counts and the Euclidean value of $<V/V_{\rm max}>$. 
The source counts exhibit large differences from the observed counts.
In most cases the value of $<V/V_{\rm max}>$ differs substantially 
from the observed value of 0.34 for BATSE-based samples, with offsets
ranging from $6-24$ sigmas. Given that
the value of $<V/V_{\rm max}>$ is well established and that it
is a cosmological distance criterion, we recommend that any study
of the GRB luminosity function take its value into account.

\end{abstract}

\section{Introduction}

Since there are no large systematically selected samples of gamma-ray
bursts (GRB) with redshifts available yet, the derivation of their
luminosity function requires a luminosity criterion to estimate the
redshifts. In this
paper, we use the results of five recent studies using one or more luminosity
criteria to compare the resulting luminosity functions and to derive 
from each the expected source counts and the Euclidean value of
$<V/V_{\rm max}>$.

These studies are presented in detail in Fenimore and Ramirez-Ruiz (2000) 
(FR), Schmidt (2001) (S), Schaefer, Deng and Band (2001) (SDB), 
LLoyd-Ronning, Fryer, and Ramirez-Ruiz (2002) (LFR), and Norris (2002) (N). 
The luminosity criteria used in these studies are variability (FR, SDB,
LFR), spectral lag (SDB, N) and spectral hardness (S). References to 
original work on these luminosity criteria are given in the individual 
papers. 

Our use of the spectral hardness was based on its correlation 
with $<V/V_{\rm max}>$ (Schmidt 2001), which we interpreted
in terms of a correlation between the peak luminosity of GRBs and 
the observed photon spectral index $\alpha_{23}$ (based on BATSE 
channels 2 and 3). This  appears to be supported by the correlation of 
the isotropic-equivalent energy $E_{\rm rad}$ and $\alpha_{23}$ shown in
Figure 1. The data used are based on BeppoSAX observations of 12 GRBs
with redshifts (Amati et al. 2002). We derived the values of $\alpha_{23}$ 
from those given for the Band $\alpha$ and $\beta$ parameters
and the redshift. It is curious, however, that no correlation is seen 
between $E_{\rm rad}$ and either the Band $\alpha$ or $\beta$ parameter.

\begin{figure}
\plotfiddle{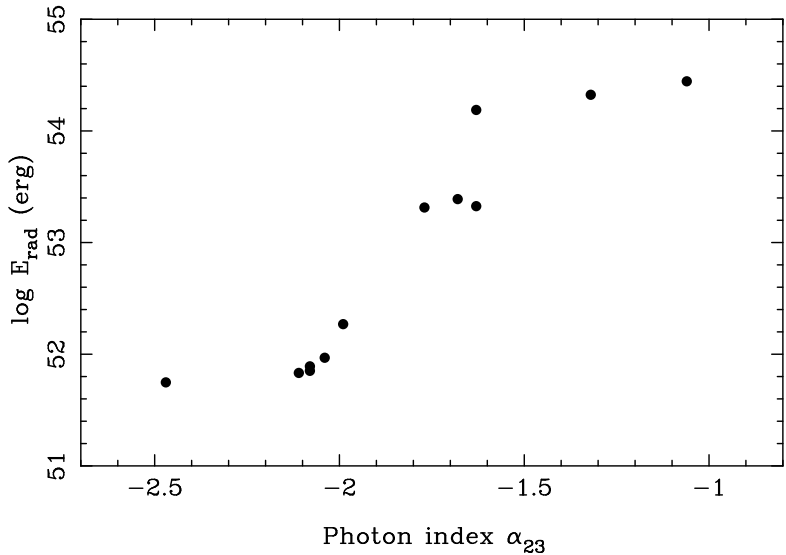}{2.6in}{-90.0}{50.0}{50.0}{-200}{270}
\caption{Isotropic-equivalent energy $E_{rad}$ versus the photon
spectral index $\alpha_{23}$ for 12 GRBs with redshifts, based
on BeppoSAX data (Amati et al. 2002). The energies were taken
directly from the Amati et al. paper. The values of $\alpha_{23}$
over the range $50-300$ keV were derived from the values
of the Band spectral parameters $\alpha$ and $\beta$ in the rest
frame, and the redhifts given in the paper.}
\end{figure}

\section{Luminosity Functions}

Generally, the authors used one or two luminosity criteria to assign 
peak luminosities to each GRB in a sample complete above a stated
flux limit. The luminosity function $\Phi(L,z)$ was then derived, see
the references. Since 'LFR' and 'N' gave only the luminosity 
distribution of their samples, we derived the luminosity function 
using the spectral properties and cosmological parameters adopted 
by the authors.

The redshift dependence or cosmological evolution of $\Phi(L,z)$ 
derived or adopted in four of the studies was a density evolution 
$\sim (1+z)^3$ up to redshifts $z = 2 - 10$. 
'LFR' concluded that in addition there was luminosity
evolution $(1+z)^3$. 'N' did not consider any evolution.
Details about the application of the luminosity criteria, and the
evolution of the luminosity function can be found in the individual 
papers. Finally, since not all luminosity functions were given in 
terms of numbers per unit volume per unit time, we applied a 
normalization factor such that, following the procedure discussed
in the next section, each produces a rate of 53 GRBs 
per year above a photon flux of 5 ph cm$^{-2}$ s$^{-1}$.

Figure 2 shows the luminosity functions $\Phi(L,z)$ at $z=0$,
where $L$ refers to the peak luminosity $L_{peak}$ at a time
resolution of 1024 ms. The cosmological parameters used are
$\Omega_M = 0.3$, $\Omega_{\Lambda} = 0.7$, and
$H_0 = 65$ km s$^{-1}$ Mpc$^{-1}$. The 'N' luminosity function extends
to $\log L_{\rm peak} \sim 46$ where its ordinate is $\sim +2.3$.
The five luminosity functions show considerable divergence, partly
caused by differences in the cosmological evolution.

\begin{figure}
\plotfiddle{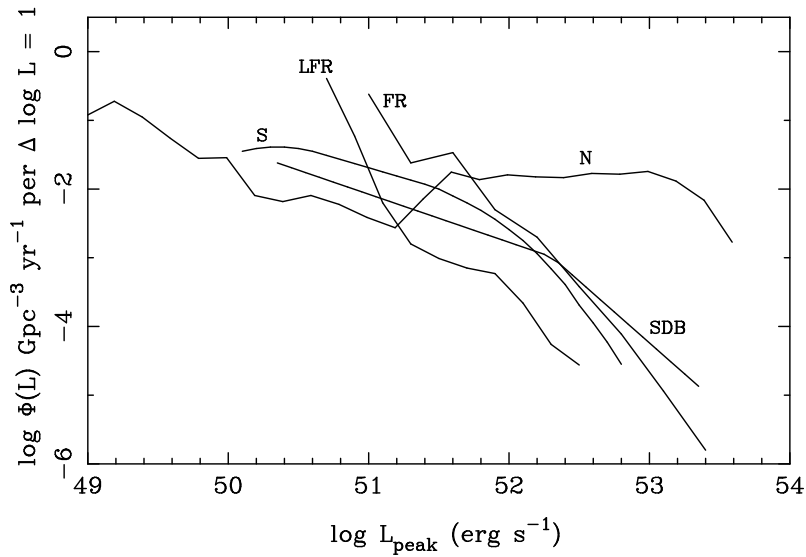}{2.6in}{-90.0}{50.0}{50.0}{-200}{270}
\caption{Luminosity functions at $z=0$ from data given in Fenimore 
and Ramirez-Ruiz (2000) (FR), Schmidt (2001) (S), Schaefer, Deng 
and Band (2001) (SDB), LLoyd-Ronning, Fryer, and Ramirez-Ruiz 
(2002) (LFR), and Norris (2002) (N). The abscis is the
isotropic-equivalent peak luminosity $L_{peak}$ over the energy
range $50-300$ keV. The redshift dependence of the luminosity
functions is described in the references.} 
\end{figure}

\section{Source Counts and $<V/V_{\rm max}>$}

For a given luminosity function $\Phi(L,z)$, the source counts 
$N(>P)$ as a function of photon flux $P$ can be derived (see
Schmidt 2001). We show in Figure 3 the resulting expected
source counts for the five luminosity functions. Also
shown are the observed counts based on the BD2 sample of GRBs
described in Schmidt (1999). Even though we had normalized the
five luminosity functions (see section 2),
the expected source counts show considerable divergence.
The counts for 'N' are very flat, probably a consequence of not 
including any cosmological evolution. At low values of $P$ the
other four predicted counts are high, only partially caused by the 
incompleteness of the counts setting in for 
$P < 0.5$ ph cm$^{-2}$ s$^{-1}$. Obviously, most of the curves
have no predictive value in estimating the number of GRBs that
may be detected by future deep missions.

The Euclidean value of $<V/V_{\rm max}>$, which to first order
characterizes the steepness of the source counts $N(>P)$, has 
the advantage that it accounts for the variation of sensitivity 
over a survey. We have derived $<V/V_{\rm max}>$ for each of the
luminosity functions, where for the individual sources
$V/V_{\rm max} = (P/P_{lim})^{-3/2}$.  We use for the limiting 
photon flux $P_{lim} = 0.25$ ph cm$^{-2}$ s$^{-1}$, which 
characterizes both the BD2 sample and the BATSE catalog. 
The results are given in Figure 3.
Conservatively assuming that the observed value $<V/V_{\rm max}>=0.34$
has an r.m.s. error of $\pm 0.01$ (Schmidt 2001), we see that four 
of the predicted values deviate by $6-24$ sigmas. The agreement 
for 'S' reflects the fact that $<V/V_{\rm max}>$ for each spectral
class was used directly to set its contribution to the luminosity 
function.

\begin{figure}
\plotfiddle{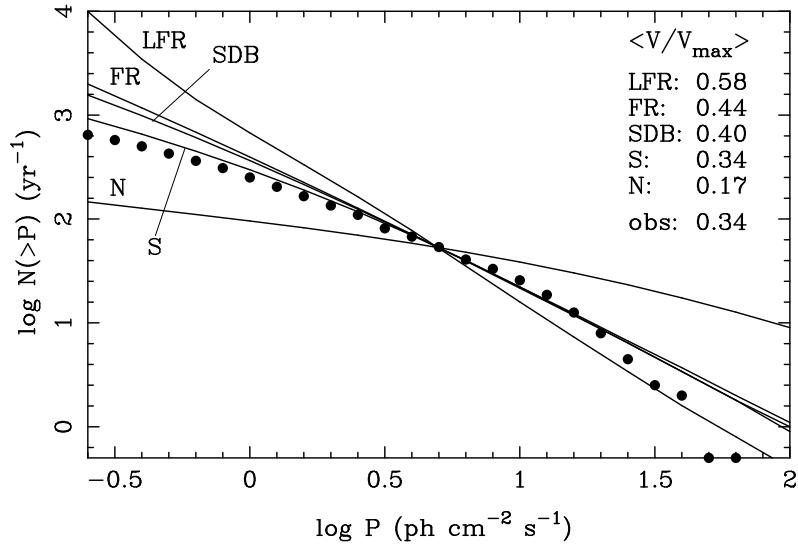}{2.6in}{-90.0}{50.0}{50.0}{-200}{270}
\caption{Predicted source
counts for five studies of the GRB luminosity function. The
photon flux $P$ is for the energy range $50-300$ keV. See
Fig. 2 for the letter code and references.
The dots represent observed counts based on the
BD2 sample (Schmidt 1999, 2001). The Euclidean $<V/V_{\rm max}>$ 
values are for $P_{lim} = 0.25$ ph cm$^{-2}$ s$^{-1}$.}
\end{figure}

\section{Discussion}

The key to understanding the systematics of Figure 3 is twofold.
First, the Euclidean value of $<V/V_{\rm max}>$ is a distance 
indicator for uniformly distributed cosmological objects (Schmidt 2001).
Second, $<V/V_{\rm max}>$ is larger for objects with stronger
positive cosmological evolution. Hence, if the expected value of 
$<V/V_{\rm max}>$ is larger than the observed one, the model 
luminosities $L$ are too low and/or the cosmological evolution is 
too strong (see Schmidt 1999, Table 1). Since the observed value 
of $<V/V_{\rm max}>$ is known quite accurately, important information 
about $\Phi(L,z)$ is missed if it is not used. We recommend that 
any study of $\Phi(L,z)$ take into account the Euclidean value 
of $<V/V_{\rm max}>$.

I thank N. Lloyd-Ronning, J. Norris, and B. Schaefer for 
useful information and discussions.

\end{document}